\documentclass{article}
\usepackage{amssymb}

\input{epsf}
\newcommand{\be}{\begin{equation}}
\newcommand{\ee}{\end{equation}}

\begin{document}

\begin{flushright}
%Liverpool Preprint: LTH 444\\
% hep-lat/9811???\\
% 3 Nov 1998\\
 \end{flushright}
  
\vspace{5mm}
\begin{center}
{\LARGE\bf Hadrons with a heavy colour-adjoint particle.}\\[10mm] 
{\large\it UKQCD Collaboration}\\[3mm]
 
{\bf  M. Foster and C. Michael}\\
\
{Theoretical Physics Division, Department of Mathematical Sciences, 
University of Liverpool, Liverpool, L69 3BX, U.K.}\\[2mm]

\end{center}

\begin{abstract}

We discuss the spectrum of hadrons with a heavy colour-adjoint particle
- motivated by the gluino of supersymmetry. Using the lattice approach, 
we explore in detail the gluonic bound states -  the `glueballino' or
`gluelump'. We also make a first determination of the  spectrum of the
`adjoint mesons' - which have a light quark and antiquark  bound to the
heavy adjoint particle. A comparison of the spectra of these  two
systems is also made.

\end{abstract}

\section{Introduction}

It is possible to explore the bound states in QCD of a particle with
adjoint  colour. This is of interest for comparison with
phenomenological models. For a pioneering study see ref.~\cite{bag}
which used  the MIT bag model.  It may also be of relevance to
experiment should a massive gluino  ($\tilde{g}$) exist  which is
sufficiently stable to form hadronic bound states. These colour-singlet
hadrons  containing a gluino have been called
`R-hadrons'~\cite{rhadron}. They include the  bound states of a gluino
and  gluons,  $g\tilde{g}$ referred to as a  `glueballino'.  Another
possibility is an R-meson, a $\tilde{g}q \bar{q}$ system, which might
also be referred to as the `hybridino' from its relationship to  the $g
q\bar{q}$ hybrid meson. The R-baryon is a $\tilde{g}q q q$ system and it
is possible that  the $\tilde{g} u d s $ state might be the lightest  of
the R-hadrons~\cite{rbaryon}. It has been proposed~\cite{cosmo} that 
these R-hadrons may have astrophysical significance as components of 
cosmic rays and,  in this case, the  mass differences between different
R-hadrons play a crucial role in  determining whether the appropriate
states could be stable.

A non-perturbative study of these states from first principles is
possible  by using numerical lattice techniques.  In  this case, it is
convenient to treat the  gluino in the heavy-gluino limit. This will be
appropriate if the  gluino mass is large compared to the QCD scale of
order 1 GeV. In this  limit, the fermionic nature of the gluino will be
irrelevant and one can use  a static adjoint source. In this context,
the gluonic bound states are known as the  `gluelump'~\cite{gl},
while we choose to call the  quark-antiquark bound states the `adjoint
meson'.  We will be unable to address the issue of the  spectrum of
`adjoint baryons'.

 We use quenched lattices to explore the spectrum of the gluelump. This has 
been studied previously~\cite{gl} but only limited results exist 
for SU(3) colour~\cite{su3adj}. A preliminary version of our work 
has been presented elsewhere~\cite{msf}. Here we make a thorough study 
of many $J^{PC}$ states and we extract the continuum limit of the mass 
differences between the lower lying states. 

 The adjoint mesons have not been studied previously on a lattice. One 
reason is that, because the heavy adjoint particle does not propagate
spatially, the light  quark propagators are only needed at the same
spatial sink  as source. This feature, common to studies of B-mesons and
the $\Lambda_b$  baryon, means that conventional light quark propagator
methods are very  inefficient. A promising new method allows the
relevant light quark  propagators to be evaluated from nearly all
sources to nearly all sites~\cite{stoc}. This has been used successfully
for static quarks and here we use similar methods to  tackle static
adjoint particles.
 Our study is exploratory and we will not be able to remove completely
the systematic  errors associated with lattice methods: extrapolation to
light quarks, continuum  limit extrapolation, etc. We also use quenched
lattices which inherently  implies at least a 10\% systematic error from
setting the scale.
  
\section{The Gluelump Spectrum}

We explore here bound states of the static adjoint source in the
presence  of the gluonic degrees of freedom. This has been explored
previously  in lattice studies~\cite{gl,su3adj,adjbreak} and we use
similar techniques.

For a heavy gluino of zero velocity, one can ignore the gluino spin  and
treat the propagation in the time direction as a product of  adjoint
gauge links. This approach, as is the case for heavy quarks in the
static limit, trades a dependence on the heavy particle mass for a 
lattice self-energy which diverges as the lattice spacing $a$ is taken
to zero. Thus we will only be able to compare with physical predictions 
for the difference of masses between states with the same adjoint
particle content. This  is, however, entirely sufficient for our
purposes.  

For the adjoint gauge links, we use the real $8 \times 8$ adjoint
matrices related to their fundamental counterparts by

 \begin{equation} 
 U_4^{{\rm Adj \ }\alpha \beta}={1 \over 2} {\rm Tr}(U_4
{\lambda^\alpha}U_4^{\dagger}{\lambda^\beta}) 
 \end{equation}

 \noindent where the $\lambda$-matrices are the conventional ones. For 
propagation of the static adjoint source, we need  the time-directed product 
of these links.
 \be
   G^A= \prod U_4^{\rm Adj}
 \ee

\begin{figure}[htb]
\epsfysize=1.5in
\centerline{\epsfbox{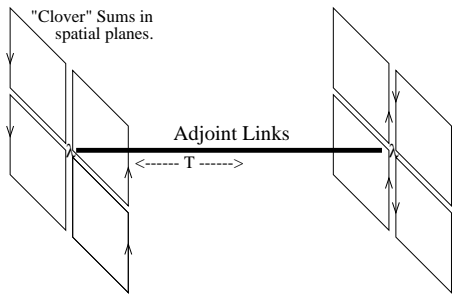}}
\caption{The gauge links involved in a  gluelump correlation.}
\label{fig:dumb}
\end{figure}

To create and destroy the gluelump states we use products of fundamental
 gauge links ($U_G$) which start and finish on the adjoint source site.
The schematic  method is illustrated in fig.~\ref{fig:dumb}. We choose
operators that are in  irreducible representations  of $O_h$  to explore
the spin structure of the gluelump states.  In the continuum limit,
states of the gluonic field are labelled by $J^{PC}$. We relate continuum
spins to those obtained from the $O_h$ subgroup by  subduction.  Note
that the bound states of an actual gluino (the glueballino) will be
fermions,  but in the limit of a heavy gluino, the gluino spin is
uncoupled so that our study gives all the relevant information.

 For speed of computation, we chose to build the gluonic operators out
of square elements.  From these we construct `clover like' operators
from various sums of these squares, projected  onto the adjoint
representation with definite charge conjugation:

 \be  
 H^\alpha = {\rm Tr} (\lambda^{\alpha} U_G \pm  \lambda^{\alpha}
U_G^{\dagger}) 
 \ee

\noindent The coefficients for creating given $O_h$ representations can
be  determined from the projection table given in~\cite{oh}. In choosing
the planar square as the building block,  we are only able to access 10
out of a possible 20 $O^{PC}$ representations. However, we expect the
lower energy  states to be created by such planar constructs.

 The correlation of interest is then given by evaluating $H^{\alpha}
G^A_{\alpha \beta} H^{\beta}$.
Diagrammatically the correlation in a typical group representation looks
like that in fig.~\ref{fig:dumb}.

Measurement of objects containing static propagators is hampered by
cumulative statistical errors from multiplying links, each with a
variance $ O(1) $.  Adjoint links are even more sensitive to this
effect~\cite{cmadj}. In order to make effective measurements at larger
times where the excited state contributions are minimised, we employ a
link integration or multi-hit technique~\cite{link} which involves
summing time-oriented links over a set of independently generated
alternatives provided by performing a local heatbath algorithm on them.
The force term is generated by the surrounding gauge links or `staples'
from the original gauge configuration. We choose to use 10 or 15 samples
of the time-directed link with this force  separated by 3
Cabibbo-Marinari SU(2) subgroup updates and then construct the average
of the adjoint links obtained  from each of them.

We measure correlations for a given state using four operators at both
source and sink.  These were constructed using two sizes of square
built from products of fuzzed links with the fuzzing algorithm performed
to two different numbers of iterations. Each iteration  of the fuzzing
algorithm makes a gauge invariant replacement of a link, $U_\mu(x)$, 
according to a sum over 4 staples:

 \be	U_\mu(x) \rightarrow P_{SU(3)}  \left( C U_\mu(x) + \sum
U_\nu(x) U_\mu(x+\hat{\nu}) U_\nu^\dagger(x+\hat{\mu}) \right)
~\label{eq:fuzz} 
 \ee 
 \noindent  where $P_{SU(3)}$ is a projection into the SU(3) group. The
fuzzing parameters,  and square sizes were tuned according to the
lattice spacing to give the best signal. We measured correlations from
all sites and time planes on various quenched  lattices, as shown in
Table~\ref{tab:lattices}. The interpolation of $r_0$ values we used is
also  given for completness.

We  then  employed the variational technique on the $4 \times 4$ matrix
of correlations in order to determine the linear combination of
operators which maximises the ground state contribution.  In practice,
since statistical errors increase with time separation $t$, we
determined the basis for the ground state from a moderate  $t$
separation ($t=1$ to $t=0$) and then explored the $t$-dependence of that
combination  to larger $t$. Since the effective mass should decrease
monotonically  with increasing $t$ to the ground state mass, we seek to
find the level of the  plateau. Since the statistical error increases
with $t$,  a sensible prescription is to select the mass from the $t$
values beyond which  the data are consistent with such a plateau. 
  We are also able to obtain estimates of the
energies  of excited states from the variational approach.

\begin{table}[h]
\caption{Lattices used in the gluelump calculation}
\label{tab:lattices}
\begin{center}
\begin{tabular}{ccccccc}
&&\\ \hline
& & \\
  $\beta$  &   Size  &   Number & Square sizes & $C$ & Fuzzing iterations&
      $r_0$ \\
          &       &   of lattices  &  & & \\
&&\\ \hline
&& \\
%  5.7  &   $8^3 \times 16$  &   20  &1,2&4.0&10,20\\
  5.7  &   $12^3 \times 24$ &   99  &1,2&4.0&10,20&2.940\\
%  5.9  &   $12^3 \times 24$ &   20  &1,2&4.0& 8,18\\
  6.0  &   $16^3 \times 48$ &   202 &2,3&4.0&20,30 &5.272\\
  6.2  &   $24^3 \times 48$ &   60  &2,3&4.0&30,45&7.319\\
&&\\ \hline
\end{tabular}
\end{center}
\end{table}

In Table~\ref{tab:glefmass} we present the lattice effective masses from
adjacent $t$-values in the optimum variational basis,  where such
determination was  statistically significant. We also give some results
for the  first excited states.
 We compare our results with an earlier exploratory calculation of the
gluelump spectrum ~\cite{su3adj} based on 50 $\beta=5.7$ lattices. The
measurement of the $T_1^{+-}$ and $T_1^{--}$ masses were given as
2.046(89) and 2.096(89)  at $t=3:2$ respectively. These older results
are seen to have the ordering we find but to underestimate the mass
splitting.

\begin{table}[ph]
\begin{center}
 \caption{Gluelump masses (ground state and first excited state in some
cases) for different $O_h$ representations in lattice units.}
 \begin{tabular}{ccllll}

\label{tab:glefmass}
 &&&&\\
 $O^{PC}$ & $\beta$ & $t=2:1$ & $t=3:2$ & $t=4:3$ & $t=5:4$ \\
\hline \\
$T_1^{+-}$
    &5.7     &1.845(6)  &1.813(15)  &1.811(48)&1633(169)  \\
    &        &2.506(1) &2.354(92) && \\
    &6.0   &1.339(3)   &1.329(5)   &1.326(5)& 1.330(8)  \\
    &      & 1.736(3)& 1.690(6)& 1.683(17)&1.650(53) \\
    &6.2   &1.152(2)  & 1.142(3)  &1.145(3) & 1.146(9)  \\
    &      & 1.469(3)& 1.445(7)& 1.446(11)& 1.429(27)\\
$T_1^{--}$
     &5.7       &2.101(8)  &2.006(33)   &2.078(123)&\\
     &          & 2.709(37)& 2.284(201)&& \\
     &6.0     &1.505(2)   &1.486(4)   &1.486(8) &1.495(21)  \\
     &        &1.883(4) &1.829(10)&1.779(28)&1.674(93)\\
     &6.2     &1.276(3)  &1.261(5)   &1.247(8)  &1.249(12) \\
     &        &1.587(4)&1.534(7)&1.532(21)&1.446(56) \\
$T_2^{--}$
    &5.7       &2.242(10)   &2.280(45)   & 2.155(310) &  \\
    &          &2.740(29)&&&\\
    &6.0     &1.593(2)   &1.579(4)   &1.549(10) & 1.557(30)  \\
    &        & 1.946(4) &1.892(10) & 1.900(42)&1.874(145) \\
    &6.2     &1.347(3)  &1.331(7)   &1.322(14) &1.296(17)  \\
    &        &1.646(4) & 1.609(8) & 1.570 (26) & 1.502(107) \\
$E^{+-}$
     &5.7       &2.470(18)   &2.230(80)   &1.862(465) &  \\
    &6.0     &1.759(3)   &1.735(8)   &1.744(22) &1.683(78)  \\
    &6.2     &1.469(4)  &1.452(9)   &1.451(23) & 1.426(41)  \\
$A_2^{+-}$
    &5.7       &2.542(26)   &3.023(347)      &  & \\
    &6.0     &1.779(4)   &1.775(13)   &1.684(33)&1.667(114)   \\
    &6.2     &1.499(6)  &1.477(9)   &1.460(28) &1.399(43) \\
$A_1^{++}$
    &5.7       &2.628(47)   &     & &  \\
    &6.0     &1.786(5)   &1.762(16)   &1.721(48)& 1.836(157)   \\
    &6.2     &1.502(7)  &1.457(15)   &1.474(41) &1.539(92)  \\
$E^{++}$
    &5.7       &2.897(50)   &   &    &  \\
    &6.0     &1.936(6)   &1.883(16)   &1.927(60) &  \\
    &6.2     &1.603(6)  &1.575(14)   &1.563(28) &1.556(91)  \\
$T_1^{-+}$
    &5.7       &2.900(37)   &      &  & \\
    &6.0     &1.966(4)   &1.918(14)   &1.918(45)&   \\
    &6.2     &1.641(5)  &1.613(11)   &1.599(34)&1.488(45)   \\
$T_2^{-+}$
    &5.7       &3.131(49)   &   &    &  \\
    &6.0     &2.085(5)   &2.053(18)   &2.216(104)&   \\
    &6.2     &1.727(5)  &1.698(18)   &1.683(46)&1.661(111)   \\
$T_2^{++}$
    &5.7       & 3.144(52)  &    &  &   \\
    &6.0     &2.148(5)   &2.130(18)   &2.142(98) &  \\
    &6.2     &1.788(6)  &1.700(20)   &1.749(81) &  \\
\end{tabular}
\end{center}
\end{table}

\begin{figure}[tbh]
\epsfysize=3.0in
\centerline{\epsfbox{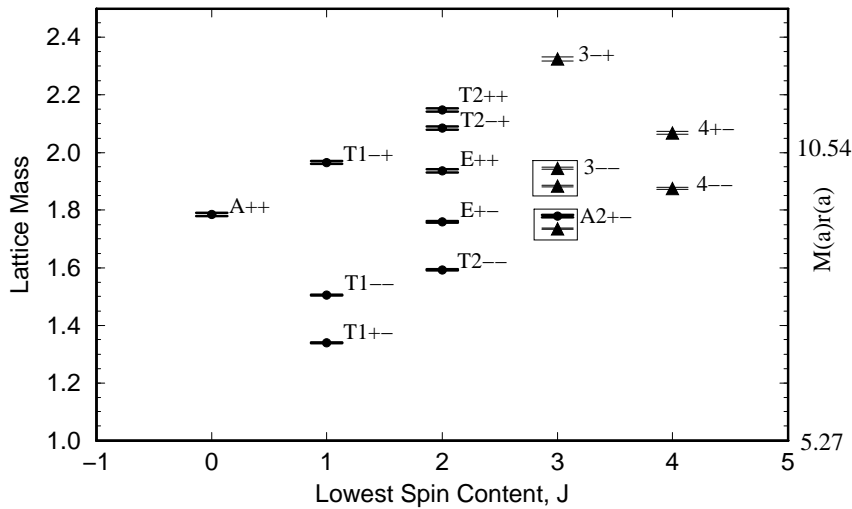}}
\caption{Gluelump Spectrum at $\beta=6.0$}
\label{fig:spec}
\end{figure}

Figure~\ref{fig:spec} shows the spectrum of states calculated at
$\beta=6.0$. The points  marked with circles are the lowest eigenvalues
from the ten measured representations.  They are plotted against $J$
assuming the lowest spin contained in  the $O_h$ representation. 
 We also determine some higher energy eigenvalues for each $O_h$
representation  we study. These could either be radial excitations with
the lowest spin  assignment or could be in a higher spin representation
allowed by that  $O_h$ representation (for example $T_1^{+-}$ can be
$J^{PC}=1^{+-}$ or $3^{+-}$). In principle a thorough study of all
eigenvalues in all $O_h$  representations in the continuum limit will
allow the $J^{PC}$ values  to be assigned unambiguously. In the present
application, we have used a solid triangle to  show plausible
assignments of $J$ for these excited states. We see that in our
$4\times4$ basis these energy eigenvalues qualitatively agree with the
expected degeneracies in the continuum spectrum (for example a $J=3$
state  has $T_1$, $T_2$ and $A_2$ degenerate levels).

\begin{figure}[tbh]
\epsfysize=3.0in
\centerline{\epsfbox{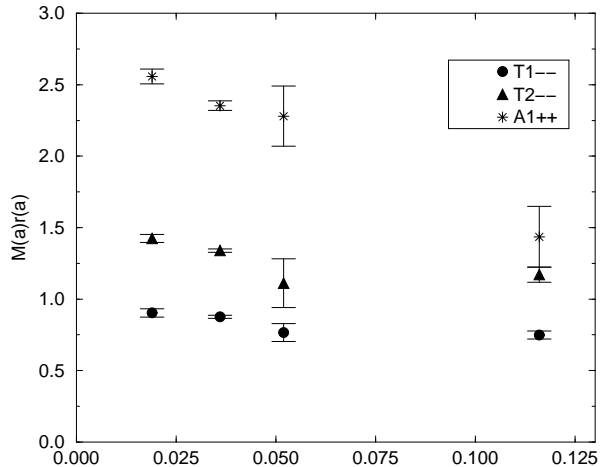}}
 \caption{The  $O^{PC}-T_1^{+-}$ gluelump mass splitting in units of
$r_0$ versus $a^2$ in units of $r_0$. The continuum limit is 
at the left and a straight line behaviour is expected for small $a$.
 }
 \label{fig:extrap}
\end{figure}

 As found previously~\cite{su3adj}, the $J^{PC}=1^{+-}$ and $1^{--}$
states are  lowest lying. Surprisingly, the lightest $0^{++}$ state is 
considerably heavier. Since the overall lattice energy contains an 
unphysical self-energy,  we examine mass differences between states for
each lattice spacing. To  determine a continuum estimate, we  study a
dimensionless quantity choosing $r_0(a)$ ($r_0$ is defined from the
force between static quarks  as $F(r_0)r_0^2=1.65$, corresponding to
about 0.5 fm, and it is measured accurately~\cite{sommer,etc} on a
lattice from the  static potential) to set the scale of the  measured
lattice mass differences $M$. Then we perform a linear fit to
$M(a)r_0(a)$, against $r_0(a)^{-2}$ since, in the  quenched
approximation, lattice corrections are of order $a^2$. The data are 
illustrated  in figure~\ref{fig:extrap} (which also shows some
additional low statistics  results from $\beta=5.9$). We thus obtain an
estimate of the continuum limit mass splittings. The  results for the
lowest few states are shown in  Table~\ref{tab:continuum} where the 
results quoted in MeV have an additional overall scale error of 10\% 
coming from the quenched approximation scale. We also show the most
plausible  assignment of the spin in the continuum limit. One rather
surprising  feature is the observed degeneracy of the $A_2^{+-}$ and
$E^{+-}$ states -  this is not compatible with a single common spin
assignment, so we have assigned  the lowest spin option in each case.

 In the continuum limit we expect the $O_h$ representations to group
into degenerate levels  with the rotational symmetry restored. Thus for
any $J$ assignment of a $E^{++}$ state (for example $J=2$ or 4), there
should be associated  a degenerate $T_2^{++}$ state in the  continuum
limit.  Our results show that the  lightest $E^{++}$ is not accompanied
by such a degenerate $T_2{++}$ state at $\beta=6.0$ or at $\beta=6.2$.
We do find, however, that  the mass difference in units of $r_0$ between
the $E^{++}$ and $T_2^{++}$ is  consistent, within statistical errors,
with decreasing (like $a^2$) to  zero in the continuum limit. This
suggests that the lattice artefact  errors (for instance those discussed
here from lack of rotational invariance) may be relatively  sizeable for
these higher lying states. This conclusion is also supported by
Figure~\ref{fig:extrap} which  shows a stronger $a^2$ dependnce for
heavier states.

\begin{table}[tbh]
\caption{Continuum Limit estimation of the  $O^{PC}-T_1^{+-}$ mass splitting.}
\label{tab:continuum}
\begin{center}
\begin{tabular}{ccccc}
 \hline
\\
State& J &$\Delta(Mr_0)_{a=0}$&Energy(MeV)&$\chi^2/{\rm dof}$\\
\\ \hline \\
$T_1^{--} $& 1&0.933(18)&368(7)  &0.874 \\
$T_2^{--} $& 2&1.438(25)&584(10)  &1.749 \\
$E^{+-}   $& 2&2.467(92)&973(36)  &0.312 \\
$A_2^{+-} $& 3&2.468(60)&972(24)  &0.113 \\
$A_1^{++} $& 0&2.771(72)&1092(28)  &0.137 \\
\\ \hline
\end{tabular}
\end{center}
\end{table}

\subsection{Potentials as $R \to 0$}

There is a correspondence between the gluelump energies we have just
determined  and the limit of excited gluonic potentials as $R \to 0$.
This has been noted  before~\cite{gl}. Here we are in a position to 
explore the consequences of this relationship more fully since 
we have determined the gluelump spectrum in detail. 

The potential between fundamental colour sources at separation $R$ has 
been widely studied. Of special interest are the gluonic excitations of 
this potential - corresponding to excited energy levels~\cite{liveu,pm,
morn}.  In the limit as $R \to 0$, the static source and anti-source
will be at the same  site and hence their colour can be combined in a
gauge invariant way - creating a  singlet  and an adjoint colour source.
 This latter is just the situation we study here:  the gluelump is an
adjoint source in the presence of a gluonic field, while the  colour
singlet correlation is given  glueball exchange (plus a vacuum
contribution when  the $J^{PC}$ representation of the object created  is
$0^{++}$).

Thus for the generalised Wilson loop in the limit of zero spatial
separation, we have 

 \be
 \lim_{R \to 0} W(R,t) = c e^{-M_{\rm gluelump }t}+c'e^{-M_{\rm glueball }t}
 \ee

\noindent In the large $t$ limit, the lighter of the two states will
dominate the  correlation function.  In most cases of present interest,
the gluelump state  is lighter than the glueball.  We can obtain a
relationship between the gluonically-excited states of the generalised
Wilson loop, in the limit $R \to 0$,  and those measured in the gluelump
spectrum. This relationship in the continuum is obtained by subducing
the SU(2) representations appropriate to the  gluelump  to the
$D_{\infty h}$ representations appropriate to the generalised Wilson 
loop when $R \ne 0$. The latter representations are labelled for
$J_z=0$, 1 ,2  as $\Sigma$, $\Pi$, $\Delta$ where $z$ is the axis of
separation of  the fundamental sources which are $R$ apart. The other
labels of the representations  are $g,\ u$ for $CP=\pm 1$ and, for the
$\Sigma$ states only, an  additional $\pm$  label indicating whether 
the sate is even/odd under  reflection in the plane containing the
$z$-axis. By subducing the irreducible representation  of the gluelump
with $J^{PC}$ we will find $D_{\infty h}$  representations with $J_z=
-J, \dots, J$; labels $g,\ u$ given by $CP$  and, for any $J_z=0$ states,
an additional label given by $P (-1)^J$.  These relationships are
tabulated in Table ~\ref{tab:sub}.

The same identities as $R \to 0$ also apply explicitly to the lattice
discretisation. Then the   $O_{h}$ representations appropriate for the
gluelump can be subduced into the  $D_{4h}$ representations appropriate 
for the  generalised Wilson loop with $R \ne 0$.  Thus, as has been
emphasised previously~\cite{gl},   the ground state gluelump with 
$1^{+-}$ ($T_1^{+-}$)  implies that as $R \to 0$ there must be a
degeneracy of the  two-dimensional  $\Pi_u$ state ($E_u$) and a
$\Sigma_u^-$ state ($A_{1u}$). 

Although the above group-theoretical identities are a good guide to the
behaviour of  the excited gluonic potentials at small $R$,  the limit as
$R \to 0$ of the excited gluonic potential is not trivial  to extract
from lattice data with $R=a,\ R=2a, \dots$. One guide is  to consider
the gluon exchange contributions perturbatively. A way to  investigate
this is to consider the self energies of the contributions:  $2E_F$ at
$R \ne 0$ and $E_A$ at $R=0$, where $F$ and $A$ label fundamental  and
adjoint colours.  Since, to lowest order, $E_A= 9 E_F /4$ for SU(3) of
colour, there will be a mismatch and one might  expect the energy to
increase as $R \to 0$ since the adjoint self-energy is larger.  Another
way to investigate this,  is to imagine that as $R \approx 0$, there is
a gluonic field in the adjoint representation,  so that the heavy quark
and anti-quark are also in an adjoint and hence will have a Coulombic
interaction energy given by $-1/8$ of the Coulombic energy  between a
quark and antiquark in the fundamental representation (which is
approximately given by $-0.25/R$  in lattice quenched studies). This
again suggests that the excited gluonic potential should  rise as $R \to
0$, here as $0.03/R$.

\begin{figure}[tbh]
\epsfysize=3.5in
\centerline{\epsfbox{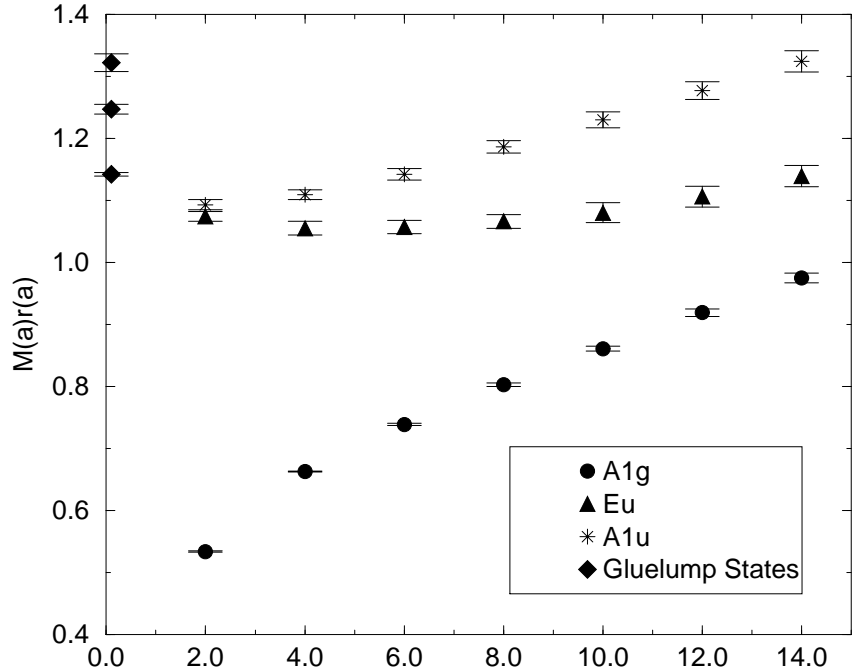}}
 \caption{The static quark potential energy in units of $r_0$ at
$\beta=6.2$ from ref.{\protect\cite{pm}} for the ground state ($A_{1g}$)
and excited gluonic  states ($E_u$ and $A_{1u}$) versus separation $R$
in lattice units. The leftmost points are (in increasing energy) the
gluelump energies for  $T_1^{+-}$, $T_1^{--}$ and $T_2^{--}$
representations. This illustrates the expected  degeneracy at $R=0$
between the two excited gluonic states which must both equal the lowest
gluelump energy.
 }
 \label{fig:wg62}
\end{figure}

Lattice data for the $E_u$ representation for  small $R$ from SU(2)
colour studies~\cite{heleu} at $\beta=2.4$ with values of
$aV_{Eu}(R)=1.31$, 1.32 and 1.38 for $R= 3a,\ 2a$ and $a$ respectively
do  qualitatively support these estimates of the small $R$ behaviour of
the excited gluonic potential and are consistent with a limit  as $R \to
0$ which agrees with the lattice gluelump energy~\cite{adjbreak} of
$aE_{\rm gluelump}= 1.50$. Also in fig.~\ref{fig:wg62}, we show the
comparison of the small $R$ excited gluonic  potentials at $\beta=6.2$
with our SU(3) gluelump analysis, where both results use the same Wilson
lattice  regularisation and so are directly related. This  figure
confirms the degeneracy of the excited gluonic energy levels as $R \to
0$ with a common  value given by the appropriate gluelump energy, as we 
found above.

 These considerations are useful~\cite{cmconf3} to understand the
extensive results on the spectrum  of excited gluonic levels that have
been determined recently~\cite{morn}.

\begin{table}[ht]

\caption{Connection between gluelump and two-body potential as $R \to 0$.}
  \label{tab:sub} \begin{center} 
 \begin{tabular}{cc} &\\ 
 \hline & \\ 
 Gluelump $J^{PC}$ &  Two-body potential states  \\  
  $1^{+-}$ &  $\Pi_u$, $\Sigma_u^-$\\
  $1^{--}$ &  $\Pi_g$, $\Sigma_g^+$\\
  $2^{--}$ &  $\Pi_g$, $\Sigma_g^-$, $\Delta_g$\\
 &\\ \hline

\end{tabular}
\end{center}
\end{table}

\section{The Adjoint-meson spectrum}

 The state with a static adjoint source bound to a quark and antiquark
is now studied. We refer to this as the adjoint-meson and label the 
states by the $J^{PC}$ of the quark anti-quark subsystem  but with a
suffix $A$ to indicate the adjoint source. In the context  where the
adjoint source is considered to be an approximation to a heavy gluino,
such a bound  state has also been called the R-meson~\cite{rhadron} and
might logically  be called a hybridino since it is the supersymmetric
partner of a hybrid meson. Note that the  bound states of a gluino will
actually be fermions, but in the limit of  a heavy gluino, the gluino
spin is irrelevant and the study with a  bosonic adjoint source gives
the required information.

The lattice adjoint-meson is generated by coupling the static adjoint
source to a light quark-antiquark system. As for the case of B meson 
studies (and those of the $\Lambda_b$ baryon), much improved statistics 
are available if one can evaluate the light quark propagators from all
sites as sources. This can be achieved using stochastic
propagators~\cite{stoc}.

 The stochastic inversion is based on the relation:
 \begin{equation}
        G_{ij} =  {\cal M}_{ij}^{-1}=\frac1Z \int {\cal D}\phi\;
        ({\cal M}_{jk}\phi_k)^\ast \phi_i\; 
        \exp \left( -\phi_i^\ast ({\cal M}^\dagger {\cal M})_{ij}
        \phi_j \right) 
 \end{equation}  
 where, in our case, ${\cal M}$ is the clover-improved 
Wilson-Dirac fermionic operator and the indices $i,j,k$ represent
simultaneously the space-time coordinates, the spinor and colour
indices.  For every gauge configuration, an ensemble of independent fields
$\phi_i$ (we use 24 following~\cite{stoc}) is generated with gaussian
probability:
 \begin{equation} 
P[\phi] =\frac1Z \exp \left(
-\phi_i^\ast ({\cal M}^\dagger {\cal M})_{ij} \phi_j \right) 
 \end{equation} 
All light propagators are computed as averages over the
pseudo-fermionic samples:
\begin{equation}
        G_{ij} = 
        \left\{\begin{array}{l}
               \langle ({\cal M}\phi)_j^\ast \phi_i \rangle \\
               \mathrm{or} \\
               \gamma_5 \langle \phi_j^\ast ({\cal M}\phi)_i \rangle\gamma_5 
               \end{array}
        \right. \label{eq:stock} 
 \end{equation} 
 where the two expressions are related by $G_{ij} = \gamma_5
G_{ji}^\dagger \gamma_5$.  Moreover, the maximal variance reduction
method is applied in order to minimise the statistical
noise~\cite{stoc}. The maximal variance reduction method involves
dividing the lattice  into two boxes ($0<t<T/2$ and $T/2<t<T$) and
solving the equation of motion numerically within each box, keeping the
pseudo-fermion field $\phi$ on the boundary fixed.  According to the
maximal variance reduction method, the fields which enter the
correlation functions must be either the original fields $\phi$ or
solutions of the equation of motion in disconnected regions.  The
stochastic propagator is therefore defined from each point in one box to
every point in the other box or on the boundary.  Hadronic correlators
are then evaluated with the hadron source and sink in different  boxes.
In order to implement this requirement, we only evaluate  correlators
for $t \ge 2$. For further details, see  see Michael and
Peisa~\cite{stoc}, especially their application to the  $\Lambda_b$
meson.  Note that, in any method which involves solving the lattice
Dirac equation, it  is not consistent to use  multihit improvement for
the time-directed gauge links.

We construct creation operators for the adjoint quark bilinear 
according to,

 \begin{equation} {\cal H}^\alpha_{\rm
Adj}=\bar{\psi}(x)\lambda^\alpha\Gamma\psi(x) 
 \end{equation}

\noindent such that the correlation function is given by combining this
with the static adjoint source $G^A$, defined previously, and replacing
quark propagator terms, $\bar{\psi}_i\psi_j$ as  $\langle\phi_j ({\cal
M}_{ik}\phi_k)^*  \rangle$ as described above. The correlation function
is given by

 \begin{eqnarray}
 \label{eq:amcor}
C(\Gamma,t_2-t_1)&=&
\sum_x \left(\bar{\psi}(x,t_1)\lambda^\alpha\Gamma\psi(x,t_1)\right)^\dagger
G^A_{\alpha\beta}(x,t_1,t_2)
\bigl(\bar{\psi}(x,t_2)\lambda^\beta\Gamma\psi(x,t_2)\bigr) \nonumber  \\
 &=&{1 \over N_s (N_s-1)} \sum_x \sum_{i \neq j} {\rm Tr}_{\rm
(fund)}\bigl( \phi_j(x,t_1)\lambda^\alpha\Gamma  ({\cal
M}\phi_i)^*(x,t_1)  \nonumber \\
 &\times&  \phi_i(x,t_2)\lambda^\beta\Gamma ({\cal
M}\phi_j)^*(x,t_2)\bigr) G^A_{\alpha\beta}(x,t_1,t_2)
 \end{eqnarray}

\noindent where $i$ and $j$ are different samples of the $N_s$
pseudofermion fields. We symmetrise the placement
of $t_1$ and $t_2$ about the  boundaries of the stochastic source at
$t=0$ or $T/2$. Where $t_2-t_1$  is odd we average over the two possible
placements.

\begin{table}[tbh]
\caption{Lattices used in the adjoint-meson calculation}
\label{tab:latrm}
\begin{center}
\begin{tabular}{cccccccc}
&&\\ \hline
& & \\
  $\beta$  &  Lattice  &  Propagator &$\kappa$ &$C_{SW}$& Gauge&$M_P$
&$M_V$  \\
  & size & samples &&&configs.&&\\
&&\\ \hline
&& \\
  5.7  &   $8^3 \times  16$ & 24 & 0.13843 &1.57& 20 & &\\
  5.7  &   $12^3 \times 24$ & 24 & 0.13843 &1.57& 20 &0.736(2)&0.938(3) \\
  5.7  &   $8^3 \times  16$ & 24 & 0.14077 &1.57& 20 & &\\
  5.7  &   $12^3 \times 24$ & 24 & 0.14077 &1.57& 20 &0.529(2)&0.815(5) \\
  6.0  &   $16^3 \times 24$ & 24 & 0.13714 &1.76& 10 &0.309(2)&0.488(5) \\
&&\\ \hline
\end{tabular}
\end{center}
\end{table}

We  measure  the correlation for observables with $\Gamma=\gamma_5,\
\gamma_i$ and $I$, corresponding to $P_A,\ V_A $ and $S_A$ (scalar)
mesons, respectively, averaging  over the components of $\gamma_i$ for
the $V_A$ case.  Other $J^{PC}$ combinations were found to be more
poorly determined with masses comparable to the scalar case, $S_A$, or
higher.

Because the stochastic inversion method evaluates so many samples  of
the correlation from each gauge configuration, it is feasible  to obtain
results from moderate numbers of gauge configurations. The  lattices
used are  detailed in  Table~\ref{tab:latrm}. At $\beta=5.7$ we use  the
 parameters for tadpole-improved clover fermions  studied
previously~\cite{shan}. With two values  of the hopping parameter, the
lighter of which corresponds approximately to the strange quark mass, 
we are able  to explore  the dependence on $\kappa$ of the spectrum and
the  extrapolation to the chiral limit.  We also evaluated stochastic
propagators at $\beta=6.0$  with  NP-improved  clover fermions to be
able to explore  the lattice spacing dependence. The masses  of the 
pseudoscalar and  vector mesons quoted in  Table~\ref{tab:latrm} come
from previous studies~\cite{shan,par} using conventional  propagators
for these lattice parameters.

We  construct isotropic extended (fuzzed) operators  by replacing the
light quark  fields  using 

\begin{equation} 
\label{eq:phifuzz}\psi_l(x)^\mathrm{fuzzed} =
\Sigma_{\pm(\mu=1,3)}{U^{(l)}_\mu(x)\psi(x+la_\mu)}
\end{equation}

\noindent where $U_{-\mu}(x)=U^\dagger_\mu(x-\hat{\mu})$ in this
context. Here $U^{(l)}$ is a product of fuzzed links in a straight line
of length $la$, each link defined according to  Equation~\ref{eq:fuzz}.
At $\beta=5.7$  we explored two choices of fuzzing, namely 2 iterations
with $C=2.5$ and 12 iterations with $C=4.0$, each with links of length
one in eq.~\ref{eq:phifuzz}. The smaller number  of fuzzing iterations
gave a more accurate value for the correlations and this was chosen  for
the $12^3$ spatial lattice for the $V_A$ and $P_A$  studies. For
$\beta=6.0$, we used 6 iterations of fuzzing with $C=2.5$  but used 
links of length 2. By replacing in Equation~\ref{eq:amcor} with fuzzed 
operators all the $\phi$ fields at $t_1$ or  at $t_2$ or  at both ends,
we generate a $2 \times 2$ correlation matrix.

\begin{figure} [t]

\vspace{12cm} %  12 too high voff =200,0, better 180 -20
\includegraphics{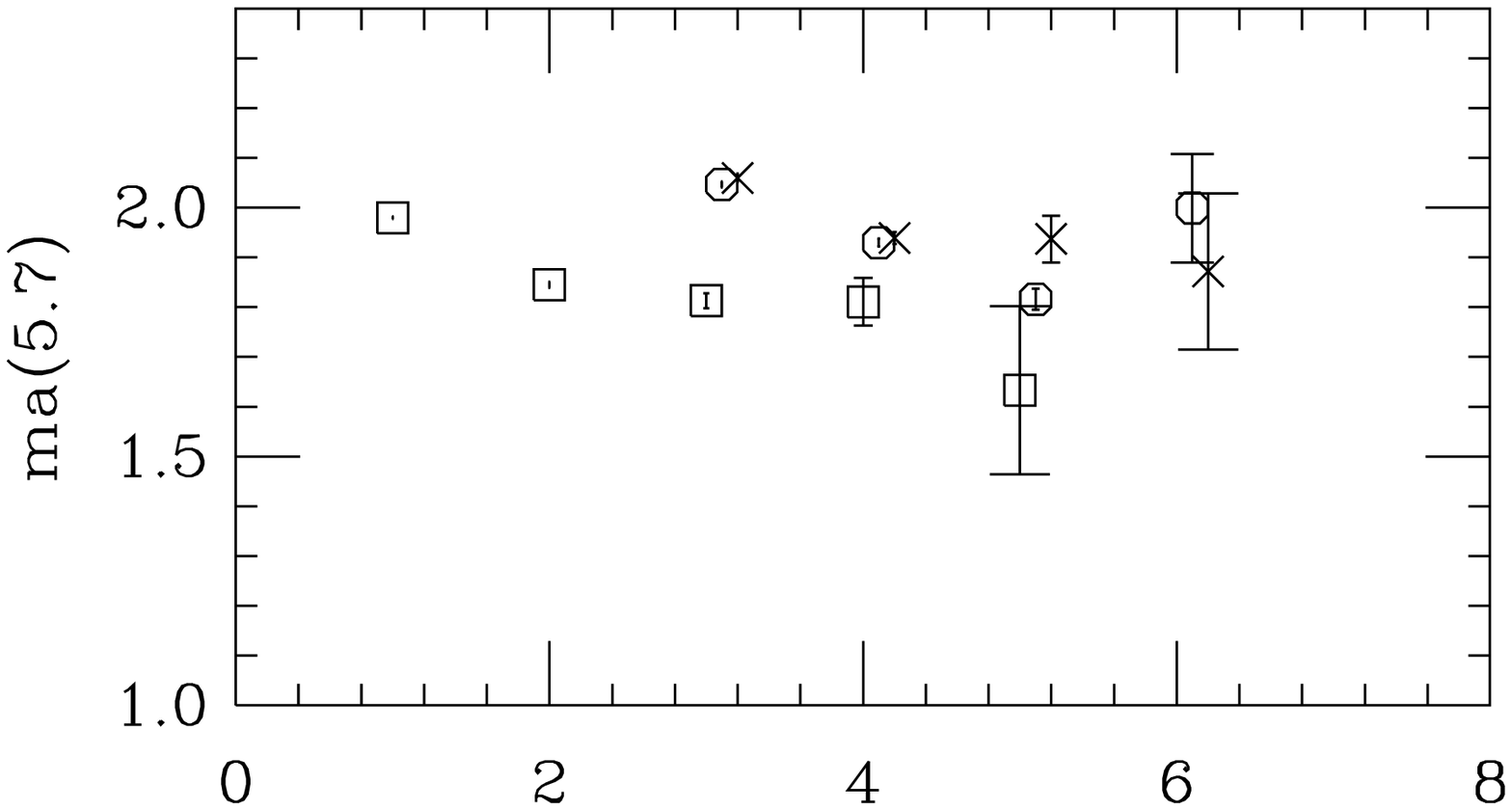}
\includegraphics{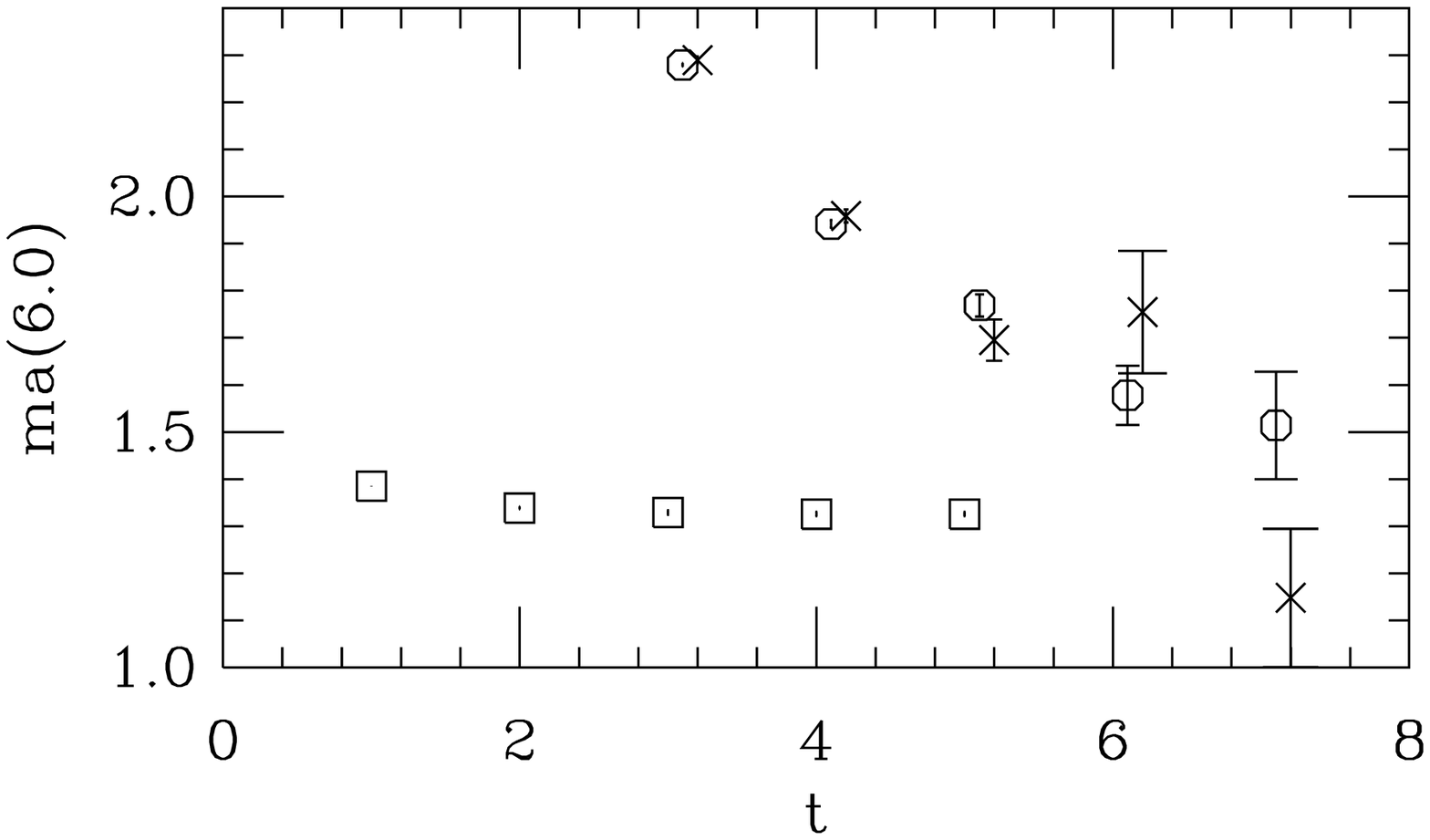}

 \caption{Effective lattice mass versus lattice time for the $V_A$
(octagon) and  $P_A$ (cross) adjoint mesons and $T_1^{+-}$ gluelump
(square) for $\beta=5.7$ (top) and $\beta=6.0$ (bottom) with light
quarks approximately corresponding  to strange.}
 \label{fig:raw}
 \end{figure}

\begin{figure} [t]
\vspace{10cm} %

\includegraphics{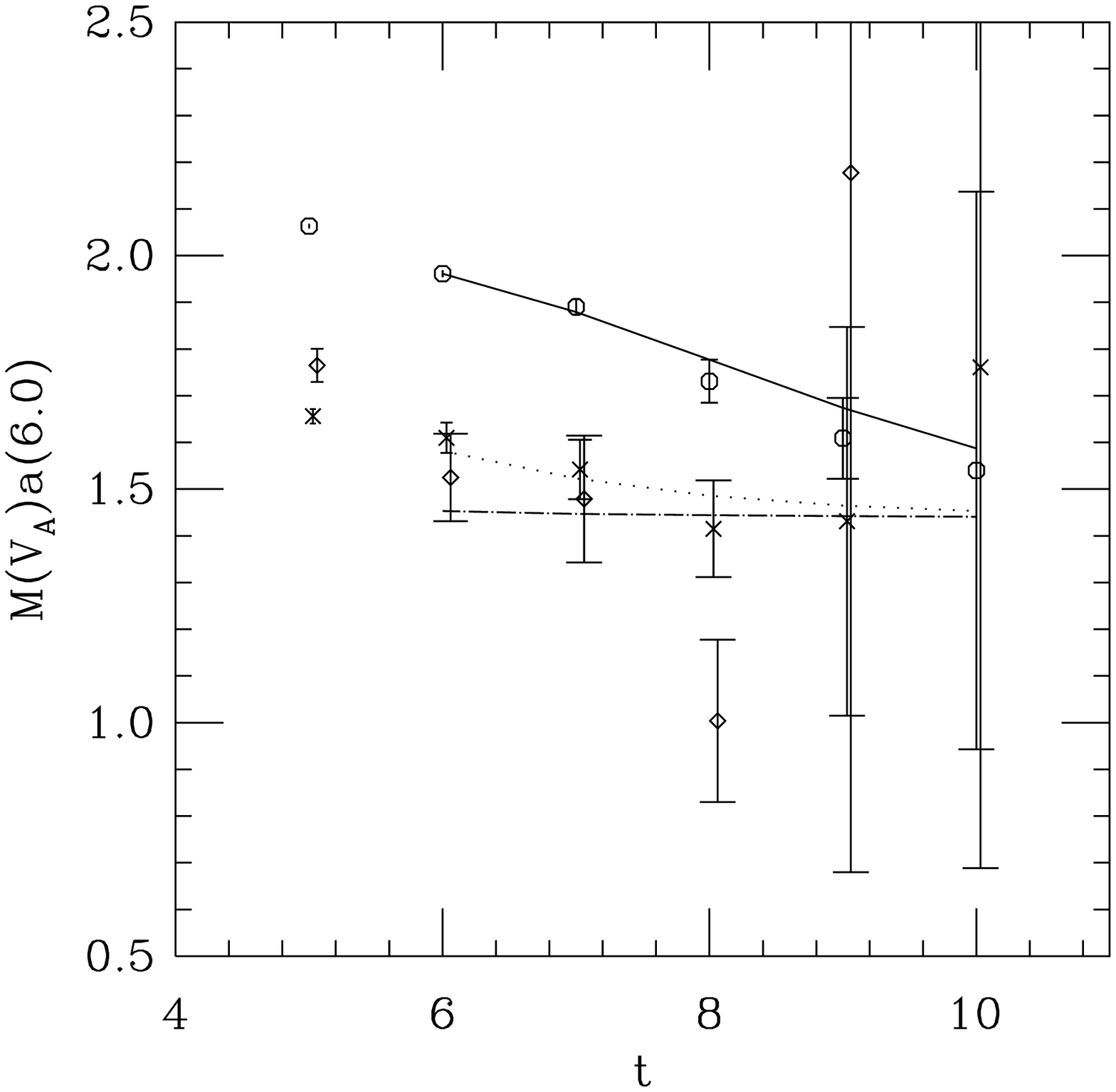}

 \caption{Effective lattice mass versus lattice time for $V_A$ at
$\beta=6.0$ with two state fit to LL(octagon), LF (cross), and FF
(diamond) correlations. Here the  light quarks approximately correspond
to strange.}
 \label{fig:fit}
 \end{figure}

As in the gluelump case, we performed a variational analysis on the data
to obtain estimates of the ground state mass. Because the excited state
contributions are relatively large, we use $t$-values of 3 and 4
 %at $\beta=5.7$ and of 3 and 4 at $\beta=6.0$ 
 to establish the optimal variational basis. The variational estimates
of the mass  are given in  Table~\ref{tab:ammass}  from the effective
mass at the adjacent  $t$-values at which the plateau is first seen
(mostly this is $t$ of 5 and 4). Some of the variational masses versus
$t$  are also shown in Figure~\ref{fig:raw},

For the larger lattices, $12^3 \times 24$ and bigger, a two exponential
fit was made to all elements of the  correlation matrix expressed as

\begin{equation}
C_{ij}(t)= \sum_{k=1,2} c_{i}^{(k)} e^{-M_k t} c_{j}^{(k)} 
\end{equation}

\noindent  between source and sink operators, $i$ and $j$. One of these
fits is illustrated in Figure~\ref{fig:fit}. Errors on the estimates of
the six parameters in the fit were made by bootstrap resampling methods
using 99 resamples of the configurations. Because of the relatively
small number of gauge  configurations, this error estimate may be
underestimated. We find the $V_A$ mass is more  accurately determined
presumably because it is  taken as the average of three  spin
components.  The  fitted masses are presented in 
Table~\ref{tab:ammass}.

\begin{table}
\caption{Ground state adjoint meson masses}
\label{tab:ammass}
\begin{center}
\begin{tabular}{ccclllll}
\\ \hline
 \\
&&&&Lattice masses:&\\
  $\Gamma$&$\beta$  &   Size  &$\kappa$ &Var. analysis
&Fit&t-range&$\chi^2/{\rm dof}$   \\
 &&\\ \hline
&&\\ 
$\gamma_5 \ (P_A)$ &&      &          &          &&&\\
&5.7&   $8^3 \times  16$   &   0.13843&1.892(57) &&&\\
&5.7  &   $12^3 \times 24$ &   0.13843&1.924(22)&1.923(27)&$4\to8$&0.8\\
&5.7  &   $8^3 \times  16$ &   0.14077&1.889(40) & &&  \\
&5.7  &   $12^3 \times 24$ &   0.14077 &1.937(47) &1.883(63)&$4\to7$&0.6\\
&6.0  &   $16^3 \times 24$ &   0.13417 &1.695(44)  &1.468(153)&
    $5 \to 10$&0.9\\\
$\gamma_i\ (V_A)$  &&&&&\\
&  5.7  &   $8^3 \times  16$ &   0.13843&1.877(19) &&& \\
&  5.7  &   $12^3 \times 24$ &   0.13843&1.926(10) & 1.925(11)&$4\to8$&0.6\\
&  5.7  &   $8^3 \times  16$ &   0.14077&1.875(24) & &&  \\
&  5.7  &   $12^3 \times 24$ &   0.14077&1.816(21) & 1.868(44)&$4\to6$&0.6\\
&6.0  &   $16^3 \times 24$ &   0.13417 &1.578(63) &1.440(70)&$5 \to 10$&1.0\\
$I\ (S_A)$&&&&&\\
& 5.7  &   $8^3 \times  16$ &   0.13843&2.262(73)&&\\
& 5.7  &   $12^3 \times 24$ &   0.13843&2.273(38)&&\\
& 5.7  &   $8^3 \times  16$ &   0.14077&2.337(173)&&\\
& 5.7  &   $12^3 \times 24$ &   0.14077&2.280(51)&& \\
&&\\ \hline
\end{tabular}
\end{center}
\end{table}

We find that the $P_A$ and $V_A$ states are lightest with the scalar
state having a weaker signal and lying significantly higher. We
concentrate in our discussions on these lower-lying  states. 

There are several systematic errors that contribute to the measurements
of the mass of these states. We are using a finite lattice volume,
finite lattice spacing and an  unphysically large quark mass and hence
there  will be extrapolation errors. There are errors in extracting the
ground state from  the large $t$ plateau also. Of course the error from
using the  quenched approximation applies too. We now discuss the 
extraction of the masses of physical significance.

The  effective masses  are plotted in Figure~\ref{fig:raw}. These masses
are generated from the optimum combination of paths found  in the
variational analysis  and are plotted as a function of lattice time. We
see that at $\beta=5.7$, a plateau is attained indicating that excited
state mass contributions are removed. Furthermore the values from the
fits  to the correlations are in agreement with this plateau value, as
shown in  Table~\ref{tab:ammass}.
 There is more excited state  contamination in the $\beta=6.0$
measurements as shown by the slower approach to a plateau. This is
related to the physical  time extent of the correlation which is
considerably shorter for a given lattice time in the $\beta=6.0$ case (a
factor of approximately 1.8). Fitting two states to  the matrix of
observables for a range $t$-values   will be a safer way to extract the
ground state mass in this case. This is seen  in Table~\ref{tab:ammass}
to give a lower mass value than the variational method  which strictly
gives an upper limit.  For the pseudoscalar case at $\beta=6.0$ the
signal is considerably more noisy so the plateau assignment is even less
clear.

 At each $\beta$ value the vector and pseudoscalar adjoint meson  masses
are very similar. As an estimate of the mass difference, we combine the
variational and fit values for strange light quarks at $\beta=5.7$ which
yields $M(P_A)-M(V_A)=50(70)$ MeV.  This result is consistent with the
expected situation,  as discussed later, that the pseudoscalar meson is
slightly heavier than the vector.   Since the vector adjoint meson mass
is better determined,  we base most of our subsequent conclusions on it.

 \begin{figure}
\vspace{8cm} %
\includegraphics{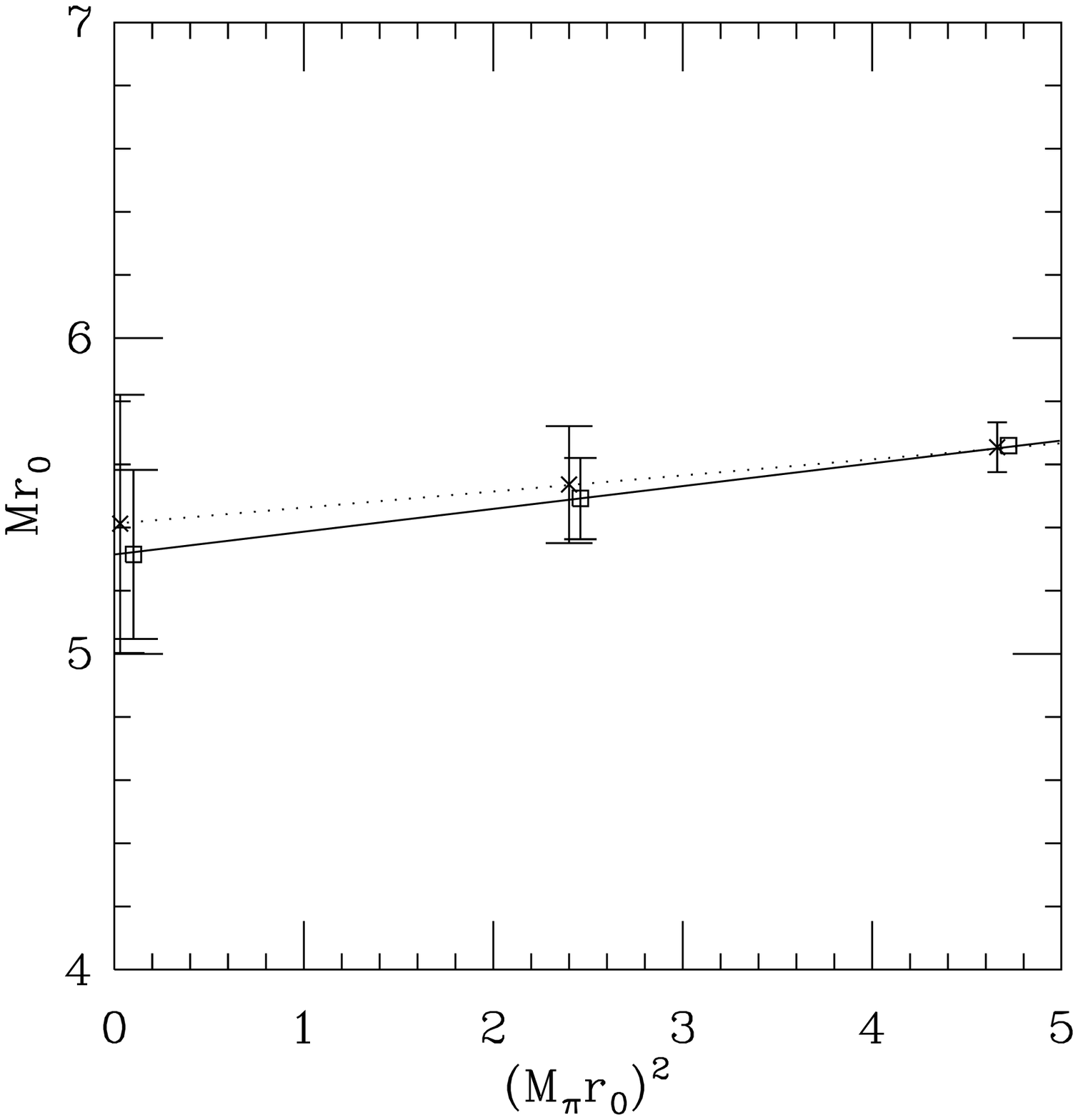}
 \caption{Dependence of the $P_A$ (crosses) and $V_A$ (squares) masses
in units  of $r_0$ on the  light quark mass (evaluated as  $r_0^2
M_{\pi}^2$)  at $\beta=5.7$, showing the extrapolation to the 
chiral limit.
 }
 \label{fig:chiral}
 \end{figure}

\begin{figure}
\vspace{8cm} %
\includegraphics{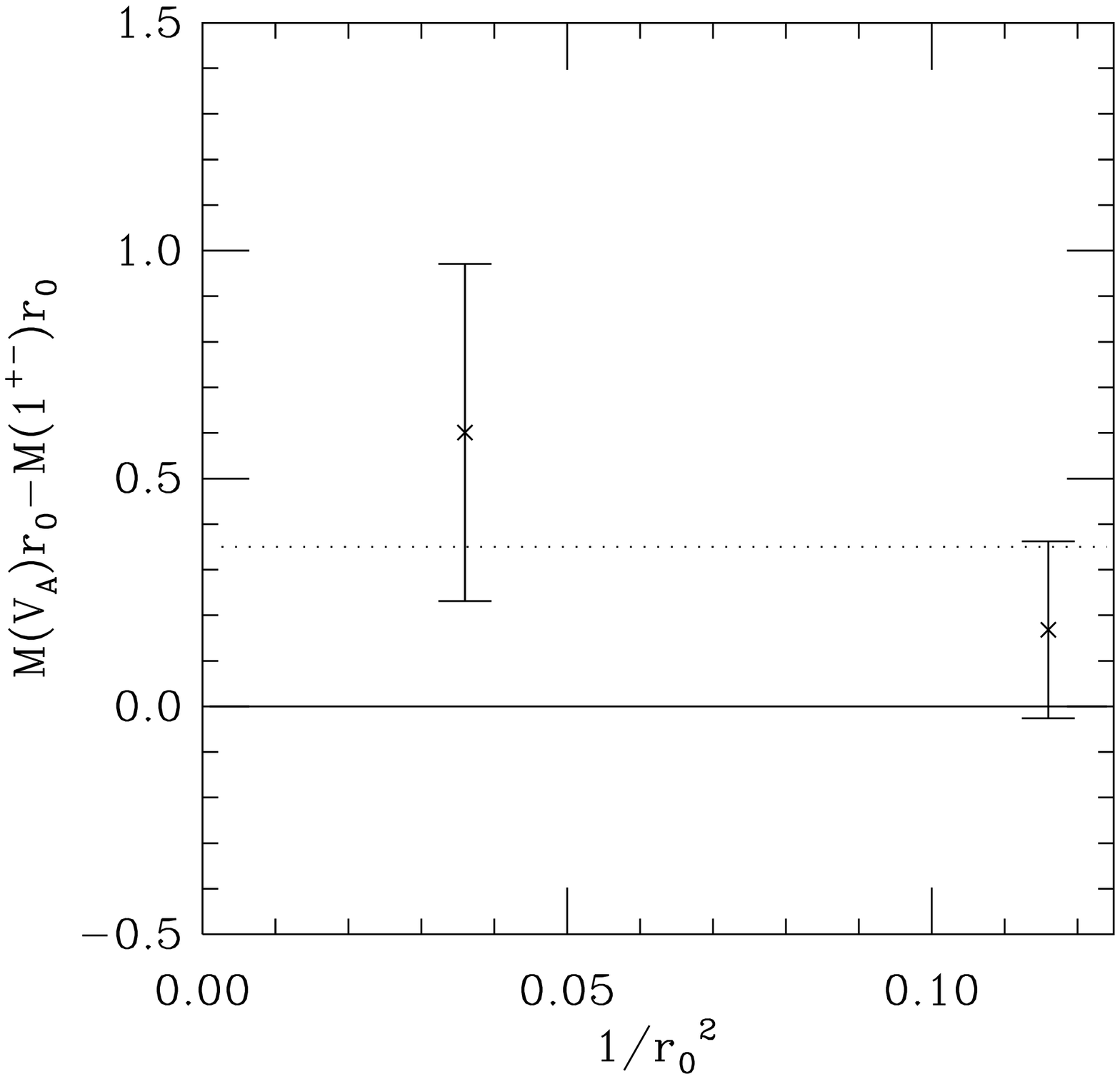}
 \caption{ The mass difference  in units of $r_0$ between the vector
adjoint meson  and the lightest gluelump state versus lattice spacing
squared ($r_0^{-2}(a) \sim a^2$). The dotted horizontal line is the
experimental pion mass (0.35 in units of $r_0$). The adjoint meson
results have approximately the same light quark mass (strange).
 }
 \label{fig:amlatdep}
\end{figure}

We now consider finite lattice volume effects. At $\beta=5.7$ we have
two lattice volumes available for direct comparisons. We see no
significant discrepancies between the adjoint meson masses on these
lattices. The spatial volume used at $\beta=6.0$ is comparable  to the
larger volume at $\beta=5.7$, so should be safe. Of course, in the 
limit as the light quark mass becomes chiral, these spatial volumes
might  be inadequate.

The $\kappa$-values used in the calculation corresponded to rather heavy
quarks (strange quark or heavier). The adjoint meson, which we model, is
composed of $u$ and $d$ quarks. At $\beta=5.7$ we used two values of the
quark mass. This enables us to make an extrapolation to light quarks as
illustrated in  Figure~\ref{fig:chiral}. We plot the  data from the fits
above as $Mr_0$ against $(M_{\pi}r_0)^2$ where we  expect both
quantities  to be approximately linear in the constituent quark mass.
Thus the figure enables us to make a linear extrapolation in the light
quark mass with the chiral limit being the point at which $(M_{\pi}
r_0)^2=0$. We find masses in lattice units at the chiral limit  of 
light quarks of $M$=1.808(91) for $V_A$ and 1.841(139) for $P_A$. Note
that this  extrapolation corresponds to a difference in mass between a
vector adjoint  meson with chiral light quarks  and one made of
$s$-quarks of 73(55)MeV. 
  
 The slope of $Mr_0$ versus $(M_{\pi}r_0)^2$ that we find at $\beta=5.7$
is consistent, within the large statistical errors,  with that
found~\cite{stoc} in similar studies of the B-meson  and $\Lambda_b$
baryon using static $b$-quarks. As noted there,  quenched lattice 
studies tend to find a smaller mass difference than  experiment.  The
experimental value of  the $B_s$ to $B_d$ mass difference is 96 MeV and
we  might expect the difference of chiral and $s$-quark adjoint mesons
to be twice this, which is a smaller value than that we found above.
Thus we may conclude that the estimate of the chiral limit  of the
adjoint mesons quoted above may well have some systematic  error, coming
either from the extrapolation or from the quenched approximation.

We now consider finite lattice spacing effects and the continuum limit. 
Since the self energy is unphysical, we study differences  between the
lightest gluelump and the adjoint meson masses at the same $\beta$ and
for a light quark mass corresponding to $s$ quarks. Thus the self energy
of the adjoint source is cancelled and we can  extract a continuum limit
of  this splitting.  The adjoint meson results use an improved SW-clover
action to reduce  order $a$ effects. In practice we have used a tadpole
improved ansatz for $C_{\rm SW}$ at $\beta=5.7$ which will not
completely remove order $a$ effects while  the $\beta=6.0$ measurement
uses a non-perturbative improvement coefficient, $C_{\rm SW}$, which
should remove O(a) effects completely.  We plot the mass differences
versus $a^2$ in Figure~\ref{fig:amlatdep}.
 This figure shows that the errors are sufficiently large that
extrapolation  to the continuum limit is not feasible. However, the
consistency of the result at our  two lattice spacings does suggest that
they  may  be  good estimates of the continuum value. Combining the 
two values then implies a difference $M_{s}(V_A)-G(1^{+-})$=120(70) MeV.

 At  $\beta=5.7$, where we are able to make a  chiral extrapolation,  we
find  that the $V_A$  and $P_A$ adjoint mesons are  -10(103) MeV  and
34(161)  MeV, respectively, heavier than the lightest ($1^{+-}$)
gluelump.  Our  result at $\beta=6.0$, though only at one light quark
mass, suggests that  these mass differences may be somewhat larger.
Indeed, combining the values from 5.7 and 6.0 for the $V_A$ with $s$
quarks (as above with $M_{s}(V_A)-G(1^{+-})=120(70)$ MeV) with that for
the difference between $s$ quarks and the chiral limit (see above:
$M_s(V_A)-M(V_A)=73(55)$ MeV) yields an overall estimate of  the mass
difference in the chiral limit of $M(V_A)-G(1^{+-})=47(90)$ MeV.

\section{Discussion}

 We have made a first non-perturbative study, albeit in the quenched
approximation,  of the adjoint meson spectrum. We are also able to
compare our results  for the adjoint meson and the gluelump, since the
unphysical lattice self-energy  cancels in this comparison. We first 
summarise some of the phenomenological predictions for these spectra. 

 In one of the first analyses of this state~\cite{bag}, Chanowitz and 
Sharpe present a bag model calculation of the adjoint-meson mass
spectra, over a range of $M_{\rm gluino}$. They find that the $J^{PC}$
ordering of the vector and pseudoscalar states places the vector 
particle as being the lighter of the states examined over the range of
$M_{\rm gluino}$ examined.  Their determination of the spectrum places
the lightest gluelump (glueballino) as slightly heavier than the
adjoint-meson for larger $M_{\rm gluino}$. Bag model calculations also 
give the $J^{PC}=1^{+-}$ gluelump as the ground state - the `magnetic'
gluon  mode.

The ordering  of the conventional meson spectrum can be understood 
qualitatively from the colour-spin interaction arising from one gluon 
exchange~\cite{gg}. This interaction makes the pseudoscalar meson 
lighter than the vector. Now for adjoint mesons, this same one gluon 
exchange will have a coefficient $-1/8$ of the conventional meson case. 
This suggests that the level ordering should be reversed - with the
vector adjoint  meson being lighter. However, the splitting would be
much reduced -  by a factor of eight. For our light quark masses, 
the $\pi$ - $\rho$ splitting is anyway smaller than experiment, 
so we expect near degeneracy of the $V_A$ and $P_A$ states - as indeed 
is consistent with our results.

 For the flavour non-singlet adjoint meson states, there will be extra 
terms in the correlation which we have not evaluated. Also there will 
be mixing with gluelumps states - especially for the vector adjoint 
meson which mixes with the relatively low-lying $1^{--}$ gluelump.

 If the gluino turns out to be the lightest supersymmetric particle and 
it is stable,   it is of interest to establish the set of hadronic bound
states of the gluino which are  stable. We can make a start on this
study by comparing the glueball  and adjoint meson spectra we have
determined in the quenched approximation. The lightest gluelump has the
gluonic field with $J^{PC}=1^{+-}$. The next gluelump state is 368(7) MeV
heavier with  $J^{PC}=1^{--}$ and,  if unmixed, will not be able to
decay hadronically to the ground  state since both $\pi$ and $2\pi$
modes are forbidden (by isospin  and parity respectively).
 The lightest non-singlet adjoint mesons are the $V_A$ and $P_A$ and we
find them to be somewhat heavier than the lightest gluelump state
$G(1^{+-})$, although with a significant systematic error  coming from
the extrapolation to light quark masses. We obtain
$M(V_A)-G(1^{+-})=47(90)$ MeV. For an  adjoint meson composed of $s$
quarks, we have  smaller errors since we do not need to  make a chiral
extrapolation: $M_s(V_A)-G(1^{+-})=120(70)$ MeV and
$M_s(P_A)-M_s(V_A)=50(70)$ MeV.

The S-wave  hadronic processes $V_A \to G(1^{+-}) + \pi$ and $G(1^{+-})
\to V_A + \pi$ are allowed when the $V_A$ is  composed of $u,\ d$ light
quarks.  It is  thus of interest to establish if the mass difference is
such as to make  either of these processes an  energetically allowed
decay. We are unable to answer  this categorically because of systematic
errors from the various extrapolations needed. However, our results do
suggest that the $V_A$ is indeed heavier than the $G(1^{+-})$ but that
the energy difference is less than $m_{\pi}$, so that $V_A$ would be
stable. For  the $P_A$ state, no one pion decay to $G(1^{+-})$ is
allowed. An allowed  process is $P_A \to V_A + \pi$ so  $P_A$ would be
unstable  if it is more than 140 MeV heavier than $V_A$ which does not
seem to be the case  from our results.
  As mentioned above, the flavour singlet adjoint mesons are even more 
difficult to study directly. Mixing with gluelump states may have a 
significant effect for them and could depress the $1^{--}$ gluelump
mass, for instance.

\section{Acknowledgements}

 We thank Janne Peisa for help in developing codes 
used for  stochastic fermions with maximal variance reduction. 
 We acknowledge the support from PPARC under grants GR/L22744 and
GR/L55056 and from the HPCI grant  GR/K41663, also support from 
HEFCE via JREI for the SGI Octane at Liverpool.

\end{document}